\documentclass[preprint]{aastex631}

\shorttitle{Formation of Li-rich and Super Li-rich RC Stars}
\shortauthors{Li et al.}
\usepackage{hyperref}
\usepackage{placeins}
\graphicspath{{./}{figures/}}

\begin{document}
\renewcommand{\thefootnote}{\roman{footnote}}
\title{Meridional Circulation. I. A Formation Channel for Lithium-rich and Super Lithium-rich Red Clump Stars}

\email{lixuefeng@ynao.ac.cn}

\author[0000-0001-9291-4261]{Xue-Feng Li}
\affiliation{Yunnan Observatories, Chinese Academy of Sciences, P.O.Box110, Kunming 650216, China}

\author[0000-0002-0349-7839]{Jian-Rong Shi}
\affiliation{CAS Key Laboratory of Optical Astronomy, National Astronomical Observatories, Beijing 100101, China}
\affiliation{University of Chinese Academy of Sciences, Beijing 100049, China}

\author[0000-0002-1424-3164]{Yan Li}
\affiliation{Yunnan Observatories, Chinese Academy of Sciences, P.O.Box110, Kunming 650216, China}
\affiliation{University of Chinese Academy of Sciences, Beijing 100049, China}
\affiliation{Key Laboratory for Structure and Evolution of Celestial Objects, Chinese   Academy of Sciences, P.O.Box110, Kunming 650216, China}
\affiliation{Center for Astronomical Mega-Science, Chinese Academy of Sciences, Beijing 100012, China}

\author[0000-0002-8609-3599]{Hong-Liang Yan}
\affiliation{CAS Key Laboratory of Optical Astronomy, National Astronomical Observatories, Beijing 100101, China}
\affiliation{University of Chinese Academy of Sciences, Beijing 100049, China}
\affiliation{Institute for Frontiers in Astronomy and Astrophysics, Beijing Normal University,  Beijing 102206, China}

\author[0000-0002-2510-6931]{Jing-Hua Zhang}
\affiliation{South-Western Institute for Astronomy Research, Yunnan University, Chenggong District, Kunming 650500, China}



\begin{abstract}
Current observations indicate that stars with higher rotation rates appear to maintain more surface lithium, and the majority of lithium-rich giants are indeed red clump stars. Hence, we investigate the mechanisms behind lithium enrichment in rotating red clump stars and the pathways to forming lithium-rich red clump stars.
Meridional circulation is prevalent in the radiative zone of rotating giants. We model its radial mixing  as a diffusion process and derive the corresponding diffusion coefficient based on its material transfer effect. Due to uncertainties in numerical calculations, we consider an average diffusion effect. Additionally, certain limiting conditions for the radial velocity of meridional circulation are incorporated. With varying input rotation velocities, we simulate the lithium evolution for red clump stars with this model.
Our results indicate that the material transfer effect due to meridional circulation can efficiently transport beryllium, produced by H burning, into the convective envelope. This meridional circulation can lead to lithium enrichment, with a maximum lithium abundance increment approaching \(3.0\,\rm dex\). Consequently, it is capable of forming both lithium-rich and super lithium-rich red clump stars. The degree of lithium enrichment exhibits a strong positive correlation with the rotation velocity, i.e., faster red clump stars show more surface lithium. Furthermore, our models indicate that lithium-rich red clump stars are relatively young (\(\sim 10^6\,\rm yr\)), which aligns with observation evidences.

\end{abstract}



\section{Introduction}\label{sect1}
Giants exhibiting lithium (Li) anomalies cannot be explained by standard theories of stellar evolution. In the giant phase, the surface Li abundance, denoted as \( A(\rm Li) \)\footnote[1]{\( A(\text{Li}) = \log[N(\text{Li})/N(\text{H})] + 12.\)}, is reduced by the inward expansion of the convective envelope \citep{1967ApJ...147..624I}. \citet{2000A&A...359..563C} demonstrated that the ceiling of Li abundances predicted by this evolutionary model is approximately \( 1.5\,\rm dex \). Consequently, the standard model is less capable of forming Li anomalous giants, i.e. so-called Li-rich giants with \( A(\rm Li) > 1.5\,\rm dex \). However, the proportion of the Li-rich giants is quite small (\( \sim 1\% \); see, e.g., \citet{2021MNRAS.505.5340M}), making it common practice to consider other non-standard processes, based on the standard model, to interpret the Li anomalous phenomena observed in giants.

Since the first report of Li-rich giants in \citet{1982ApJ...255..577W}, an increasing number of Li-rich giants have been discovered \citep[e.g.][]{2019MNRAS.484.2000D,2019ApJS..245...33G,2021A&A...651A..84M}. Some of the Li-rich giants have even been found that contain more Li than the interstellar medium of $A(\rm Li)\sim3.3\,dex$, known as super Li-rich giants \citep{2018NatAs...2..790Y, 2019MNRAS.482.3822S}, but they are much rarer \citep{2019MNRAS.484.2000D, 2022ApJ...931..136Z}. With the aid of asteroseismology \citep{2011Natur.471..608B}, it is possible to separate red clump (RC) stars from red giant branch (RGB) stars; both have the H burning shell, but the RC stars possess a burning He core. Among the Li-rich giants, the RC stars appear more frequently \citep[e.g.][]{2021MNRAS.505..642D, 2021NatAs...5...86Y}. 
Although spectral surveys have provided enough samples of Li-rich giants, for some general observation information such as $T_{\rm eff}$, log$\,g$, and [Fe/H] \citep{2021MNRAS.505.5340M}, they have not been found to be the key indicators of enriching Li. This result casts a shadow over the formation mechanism of Li-rich giants, suggesting that the Li enrichment may be a universal behavior among giants. Additionally, the analyses for stellar mass \citep{2022ApJ...931..136Z} and metal elements \citep{2020MNRAS.494.1348D} also show that there is no significant difference between Li-rich and Li-normal, $A(\rm Li)<1.5\,dex$, giants. However, some recent studies have shown that part of Li enrichment giants have a good correlation with rotation velocity, with faster rotators tending to have higher surface Li content \citep{2016A&A...587A..66D, 2021ChA&A..45...45D, 2021A&A...651A..84M,2023AJ....166...60T, 2023A&A...674A.157T, 2024ApJ...964...42S}. Furthermore, certain phenomena, such as infrared excess \citep{2015ApJ...806...86D, 2024arXiv241204624D} and stellar activity \citep{2022ApJ...940...12S, 2024A&A...688A..68R}, are also associated with different subsets of Li-rich giants.

As mentioned earlier, explaining the abnormal behavior of Li in giants requires consideration of some non-standard processes. The transport process of beryllium inside stars is usually mentioned, i.e., so-called Cameron-Fowler process \citep{1971ApJ...164..111C}. Because of the difference in the reaction temperature of Li and beryllium, to increase the abundance of surface Li, it is necessary to transport the beryllium produced by H burning inside the star to the convective envelope, where the temperature will not destroy Li produced by the \(\beta\) decay of beryllium. This process is first used to comprehend the Li behaviour of asymptotic giant branch stars that undergo the `hot bottom burning' process \citep[e.g.][]{1992ApJ...392L..71S}. For red giants, however, the `hot bottom burning' process does not occur, thus necessitating the consideration of extra mixing processes, such as thermohaline mixing \citep{2007A&A...467L..15C} and mixing driven by internal gravity waves \citep{2020ApJ...901L..18S, 2023ApJ...943..115L}, etc. For more on extra mixing, readers can refer to \citet{2024MNRAS.535.1243D}. Furthermore, various physical processes within binary systems have been widely discussed \citep[e.g.][]{1999MNRAS.308.1133S, 2016ApJ...829..127A, 2020ApJ...889...33Z}. In addition to directly increasing surface Li content through the physical processes described above, an indirect approach is to inhibit the Li depletion process predicted by the standard model \citep[e.g.][]{2024MNRAS.529.1423L}.  

Research on Li anomalous giants has been ongoing for more than 40 years, yet the field remains open and dynamic. As problems presented above, we will investigate the Li enrichment mechanism of rotating RC stars in current work. To operate the Cameron-Fowler process, extra mixing in the radiative zone between the H burning shell and the convective envelope is required. \citet{2019ApJ...880..125C} found that the tidal interaction in binary systems can spin up the host star to enhance the mixing effect inside the star, potentially leading to the formation of a Li-rich giant. However, \citet{2024A&A...690A.367C} indicated that this process is not common for Li-rich RC stars by analysing the radial velocities of 1400 giants. Circulatory motion, which universally occurrs in the interior of rotating stars, is a likely candidate mechanism. Rotating stars are subjected to centrifugal force, which can disrupt the spherical symmetry of a star, transforming it into an ellipsoid of rotation. The equipotential surfaces along the axis of rotation are more densely spaced than in the vertical direction of the axis of rotation, thus making the radiative flux along the polar direction greater than that along the equatorial plane. As a result, a large scale clockwise circulation will be formed in the meridian plane of the star due to the above thermal imbalance. In the Cowling point-source model, the streamlines for meridional circulation do not pass through the stellar core, but exist as the loops within the radiative envelope \citep[see][]{1982ApJS...49..317T}, and this model is a seemly approximation to the main sequence stars. Later on, \citet{1989ApJ...347..821C} indicated that the streamlines for meridional circulation will not penetrate into the convective envelope but will move horizontally along its base when the stellar convective envelope cannot be ignored. This scenario is sufficient to drive the Cameron-Fowler process inside the red giants. Meridional circulation can also occur in the convective zones \citep[see chapters 3 and 5 of][]{2000stro.book.....T}, and we will not consider the cases in present work. The circulation motion that exists between the H burning shell and the convective envelope can enable the transfer of elements and further change the abundance of the stellar surface, as seen in studies of red giants with abnormal surface CNO abundances \citep[e.g.][]{1979ApJ...229..624S, 1992MNRAS.256..449S, 2000MNRAS.316..395D}. \cite{1979ApJ...229..624S} investigated the transfer of CNO cycle products by meridional circulation within the above radiative zone for the RGB stars, and this  pattern requires centrifugal forces to be much smaller than gravitational forces (i.e. $\varkappa_r \ll 1$, and see $\rm Section\,\ref{sect23}$). At the same time, they indicated that effective circulation mixing would occur between the H burning shell and the convective envelope when the H burning shell crosses the chemical abundance discontinuity left by the inward extend of the convective envelope. Naturally, this pattern for meridional circulation can also be used to investigate surface Li behaviour in giants \citep[e.g.][]{1995ApJ...453L..41C, 2004ApJ...612.1081D}. \citet{2004ApJ...612.1081D} had analysed that faster rotators have higher Li abundances and explored the physical processes that spin up the RGB stars. Here, we turn our attention to the RC stars. To more accurately depict the relationship between the rotation velocity of the RC stars and their Li enrichment effect, in this work, we will treat the radial mixing of meridional circulation as diffusion process from the point of material transport, and track its Li enrichment effect under diverse rotation velocities.

The structure of this paper is as follows. Section \ref{sect2} details the model setup and describes the treatment for meridional circulation's radial mixing; subsequently, Section \ref{sect3} presents and analyses the results of the model predicted, and Section \ref{sect4} discusses the diffusion coefficient as well as the rotation velocity and Li abundance; finally, a brief summary is provided in Section \ref{sect5}.

\section{Method}\label{sect2}
\subsection{Modelling}\label{sect21}		
With the help of the Modules for Experiments in Stellar Astrophysics (MESA; version number 11701), the evolution model of rotating stars is established \citep[see][]{2011ApJS..192....3P,2013ApJS..208....4P,2019ApJS..243...10P}. Here are some basic settings. The OPAL equation of state tables \citep{2002ApJ...576.1064R} are adopted in our models. The OPAL opacity tables of \citet{1993ApJ...412..752I, 1996ApJ...464..943I} are adopted in the high-temperature region, while in the low-temperature region we select the opacity table from \citet{2005ApJ...623..585F}. The treatment for convection is based on the mixing length theory of \citet{1968pss..book.....C}, and the mixing length coefficient is 2.0. The convective boundaries are given by the Schwarzschild criterion, while the atmospheric boundary is determined by the Eddington $T-\tau$ relation. To ensure a homogeneous convective envelope, we add a powerful overshooting to the stellar surface ($f_{\rm ov}=0.80$; see \citet{2024MNRAS.529.1423L}). 

The chemical composition of the models is referred to \texttt{GS98}, and the input composition of the models is \(X=0.7345\), \(Y=0.2485\), and \(Z=0.0170\), respectively \citep[the present solar composition;][]{1998SSRv...85..161G}. In addition, we take the model mass as \(1.2\,M_{\odot}\). The nuclear reaction network option  is \texttt{pp\_extras.net} that includes 12 isotopes: $\rm ^{1,2}H$, $\rm ^{3,4}He$, $\rm ^{7}Li$, $\rm ^{7}Be$, $\rm ^{8}B$, $\rm ^{12}C$, $\rm ^{14}N$, $\rm ^{16}O$, $\rm ^{20}Ne$, and $\rm ^{24}Mg$. 

Then, we take rotation in account. The equatorial rotation velocity at the zero age main sequence (ZAMS) is given as $v_{\rm in,\,ZAMS}=10\,\rm km\,s^{-1}$ \citep[see][]{2014ApJ...780..159E}, and the rotation velocity will be significantly reduced due to the expansion of the stars during the giant stage \citep[e.g.][]{2024A&A...688A.184L}. Since observations show that some RC stars have high rotation velocities \citep[e.g.][]{2023AJ....166...60T, 2024ApJ...966..109S}, we will take input velocities of \(1,\ 10,\ 30,\ 50, \rm\ and\ 100\,\rm km\,s^{-1}\) in the target phases, i.e., $v_{\rm in,\,HeF}$ and $v_{\rm in,\,RC}$, as detailed in Section \ref{sect3}. According to the conservation of angular momentum, the RC progenitors would need to be about $\rm 500-1000\, km\,s^{-1}$ during the main sequence to maintain a high rotation velocity of $\rm 50-100\,km\,s^{-1}$ during the core He-burning phase. Such extreme velocities present challenges for low-mass main sequence stellar models. Without taking the velocity for the progenitors of the RC stars into account, however, our choice of 50 and $\rm 100\,km\,s^{-1}$ as input velocities for the RC stars is empirically motivated that is based on partial observed results. For instance, in the sample collected by \citet{2021A&A...651A..84M} and \citet{2000A&A...363..239D} there are giants with a rotation velocity ($v\,sini$) of about $\rm 100\,km\,s^{-1}$. In addition, analysis of the GALAH survey data by  \citet{2021MNRAS.505.5340M} revealed some giants with $v_{\rm broad}$ measurements exceeding $\rm 50\,km\,s^{-1}$. Similar finding have been corroborated by recent work from \citet{2024ApJ...964...42S}. Usually, the RC stars do not rotate too fast \citep[e.g.][]{2023A&A...673A.110M}, but high velocity samples have been observed. If angular momentum is conserved, then the high velocity input we prepare may cover the acceleration effect of some physical processes. Although \citet{2006ApJ...641.1087D} and \citet{2019ApJ...880..125C} explored the physical scenarios that spin up red giants, this is not part of our current work. Except for the default meridional circulation in MESA, we incorporate other instability processes associated with rotating stars into our models (see \citet{2000ApJ...528..368H} for more details), with specific settings as follows:
\[
\begin{array}{lp{0.8\linewidth}}
	\texttt{am\_nu\_visc\_factor = 0 }    \\
	\texttt{am\_D\_mix\_factor = 0.0333}    \\
	\texttt{D\_DSI\_factor = 1}   \\
	\texttt{D\_SH\_factor = 1}   \\
	\texttt{D\_SSI\_factor = 1}   \\
	\texttt{D\_GSF\_factor = 1}   \\
	\texttt{D\_ST\_factor = 1}   \\
	\texttt{D\_ES\_factor = 0}  \ !\  \rm Meridional\ Circulation\\
\end{array}
\]
We use the interstellar medium abundance as the initial Li abundance at the ZAMS, i.e., \(A(\rm Li)_{in, \,ZAMS}=3.3\,dex\) \citep{2003ApJ...586..268K}. $A(\rm Li)$ in the above models will drop to $\rm 0.52\,dex$ at the zero age of the core He burning ($Y_c \sim 0.98$). All of our models evolve from the ZAMS to the end of the core He burning ($Y_c=0.01$).

\subsection{Diffusion Coefficient} \label{sect22}
\cite{1979ApJ...229..624S} gave the mass variation of the H burning products entering the stellar convective envelope through meridional circulation
\begin{equation}\label{eq1}
	\frac{dm}{dt} = 2\, \pi\, r^2\, \rho\, v_{\rm MC}, 
\end{equation}
where $r$ is the radial location and $\rho$ is the density. $v_{\rm MC}$ is the radial component of the circulation velocity. Obviously, Equation (\ref{eq1}) also describes the radial mass change in the radiative zone.

\citet{1992MNRAS.256..449S} calculated the transfer of CNO cycle products from the point of view of material transfer, but this approach cannot fully reflect the time evolution information of surface abundance. Directly calculating how much beryllium is input into the convective envelope requires knowing the precise distribution profile of beryllium in the whole radiative zone and the evolution law with time. By convention, in the modelling calculations, we can approximate the material mixing as a diffusion process \citep[e.g.][]{1989ApJ...338..424P}. As for large scale circulation, \citet{1992A&A...253..173C} noted that, due to the horizontal turbulence in the radiative zone of rotators, the radial mixing for meridional circulation will be inhibited and can be treated as diffusion. In this work, we temporarily ignore the effect of horizontal turbulence and derive the diffusion coefficient from the macroscopic perspective of material transport. Then, in the radial direction, the mass change in the products of the H burning shell, $\Delta m$, due to meridional circulation can be expressed as
\begin{equation}\label{eq2}
	\Delta m = \rho\, D_{\rm MC}\,  \Delta t\, \Delta r.
\end{equation}
Here, $D_{\rm MC}$ is the diffusion coefficient of meridional circulation. Combining Equations (\ref{eq1}) and (\ref{eq2}), we have
\begin{equation}\label{eq3}
	D_{\rm MC} = 2\, \pi\, \frac{r^2}{\Delta r}\, v_{\rm MC}.
\end{equation}

In the simulation calculation, the variation in $r$ is represented as a difference. Therefore, $D_{\rm MC}$ should be expressed as
\begin{equation}\label{eq4}
	D_{\rm MC,\,i} = 2\, \pi\, \frac{r_{\rm i}^2}{|r_{\rm i}-r_{\rm i+1}|}\, v_{\rm MC,\,i},
\end{equation}
where subscript i denotes the stratification of stars.

In numerical calculations of stellar models, however, the value of $\Delta r$ is highly dependent on the spatial resolution of the models. The more stratified the star, the smaller $\Delta r$ is. It can be expected that when the stars are stratified enough, the diffusion coefficient of meridional circulation will exceed that of the convective mixing. To solve this problem, a proper approach is to consider the average effect of meridional circulation, that is, to solve the mean diffusion coefficient $\langle D_{\rm MC} \rangle$. Then, Equation (\ref{eq3}) has the form
\begin{equation}\label{eq5}
	\langle D_{\rm MC} \rangle= 2\, \pi\, \langle v_{\rm MC} \rangle\,\frac{r_{\rm 1}^2}{r_{\rm 1}-r_{\rm 2}}.
\end{equation}
In this case, $d\langle v_{\rm MC} \rangle / dr = 0$, so we can treat $\Delta r$ as $r_{\rm 1}-r_{\rm 2}$, with subscripts 1 and 2 representing the bottom of the convective envelope and the lower boundary of the region of action of Equation (\ref{eq5}), respectively. $r_2$ is near the outer layer of the H burning shell, and in general, $r_2$ is its upper boundary. Therefore, Equation (\ref{eq5}) only gives the diffusion coefficient between the H burning shell and the convective envelope.

Starting from the first law of thermodynamics, \citet{1974IAUS...66...20K} deduced the radial velocity of meridional circulation
\[
v_{\rm MC}=\frac{\nabla_{\rm ad}}{\delta (\nabla_{\rm ad}-\nabla)} \,\frac{L_r}{M_r}\,\frac{1}{g_r}\,\varkappa_{r}=\frac{\nabla_{\rm ad}}{\delta (\nabla_{\rm ad}-\nabla)} \,\frac{L_r\,\varkappa_{r}}{M_r}\,\frac{3}{4\,\pi\,r\,G\,\rho} 
\]
\centerline{\text{or}}
\begin{equation}\label{eq6}
v_{\rm MC}=\frac{\nabla_{\rm ad}}{\delta (\nabla_{\rm ad}-\nabla)} \,\frac{L'_r}{M'_r}\,\frac{1}{g'_r}\, \varkappa_{r}.
\end{equation}

The second expression in Equation (\ref{eq6}) is from \citet{2013sse..book.....K}. A similar form can also be seen in \citet{2000stro.book.....T} and \citet{2009pfer.book.....M}. Here, $L_r$, $M_r$, and $g_r$ have conventional meaning, and they with superscript $'$ indicate the characteristic value. \citet{2013sse..book.....K} replaced them by the total luminosity, the stellar mass, and the surface of gravity acceleration. $G$ is the gravitational constant. The temperature gradient: $\nabla_{\rm ad}=(d\,\text{ln}\, T / d\, \text{ln}\, P)_{\rm ad}$ and $\nabla= d\,\text{ln}\, T / d\, \text{ln}\, P$, and the expansion index: $\delta=-(\partial\, \text{ln}\,\rho/\partial\, \text{ln}\,T)_P$. $\varkappa_r$ is the ratio of centrifugal force to gravity and has the form
\begin{equation}\label{eq7}
	\varkappa_r=\frac{\Omega^2_r \, r^3}{G\,M_r},
\end{equation}
where $\Omega_r$ is the angular velocity at the location point $r$. Equation (\ref{eq6}) is suitable for the rigid body rotation\footnote[2]{It should be uniform rotation, actually. However, we primarily focus on the case of RC stars, which are more static compared to RGB stars, thus we employ the term `rigid body rotation'.} and requires $\varkappa_r$ to be small.
Further, \cite{1974IAUS...66...20K} gave the $\langle v_{\rm MC}\rangle$ of whole star
\begin{equation}\label{eq8}
	\langle v_{\rm MC}\rangle=\frac{L_0}{M_0\,g_0}\,\left \langle \frac{\nabla_{\rm ad}}{\delta(\nabla_{\rm ad}-\nabla)}\right\rangle \, \langle \varkappa_r \rangle.
\end{equation}
Here, subscript 0 represents the surface of a star. $L_0$, $M_0$, and $g_0$ is the total luminosity, the total mass, and the surface of gravity acceleration, respectively.

The radiative zone is in undegenerate state. For the perfect gas we have $\delta=1$. During the evolution, $\nabla_{\rm ad}/(\nabla_{\rm ad}-\nabla) \approx \ \rm constant$. Referring to the settings of \cite{1979ApJ...229..624S}, we have 
\begin{equation}\label{eq9}
	\left\langle \frac{\nabla_{\rm ad}}{\nabla_{\rm ad}-\nabla}\right\rangle =3.
\end{equation}
Similarly, if we integrate $\varkappa_r$ over the whole star, and get
\begin{equation}\label{eq10}
	\langle \varkappa_r \rangle=\frac{1}{r_0}\frac{\Omega^2_r}{G}\, \int_{0}^{r_0} \,\frac{r^3}{M_r}\,dr = \frac{\Omega^2_{r_0} \, r^3_0}{G\,M_0}.
\end{equation}
In the radiative zone, for the rigid body one have $\Omega_r=\Omega_{r_1}\approx \Omega_{r_0}$. Here, we adopt the hypothesis of rigid body rotation law for the convective envelope \citep[see, e.g.,][]{2006A&A...453..261P}. $r_0$ is the stellar raidus. Then, Equations (\ref{eq8}) and (\ref{eq5}) have following forms 
\begin{equation}\label{eq11}
	\langle v_{\rm MC}\rangle= \frac{3\, L_0\,\Omega_{r_0}^2\, r_0^5}{G^2\, M_0^3}
\end{equation}
and
\begin{equation}\label{eq12}
	\langle D_{\rm MC}\rangle= 6\,\pi\, \frac{r_{\rm 1}^2}{r_{\rm 1}-r_{\rm 2}}\, \frac{L_0\,\Omega_{r_0}^2\, r_0^5}{G^2\,M_0^3}.
\end{equation}

For the characteristic timescale of meridional circulation
\begin{equation}\label{eq13}
	\tau_{\rm MC}= \frac{r_0}{\langle v_{\rm MC}\rangle} = \frac{1}{3\,\langle \varkappa_r \rangle}\,\tau_{\rm KH},
\end{equation}
where $\tau_{\rm KH}$, $\tau_{\rm KH}=\frac{E_G}{L_0}=\frac{G\,M_0^2}{r_0\,L_0}$, is the Kelvin-Helmholtz timescale, and $E_G$ is the total gravitational potential energy. Because the convective timescale, $\tau_{\rm conv}=H_{P}/v_{\rm conv}$ ($H_P$ is the pressure scale height and $v_{\rm conv}$ is the convective velocity), is much smaller than $\tau_{\rm MC}$, we can use it to characterize the Li enrichment timescale of meridional circulation.

\subsection{Criterion for the Rigid Body Rotation}\label{sect23}
Equation (\ref{eq12}) can be adopted if the radiative zone between the H burning shell and the convective envelope is considered as a rigid body. This demands $\varkappa_r \ll 1$, and it can be represented as 
\begin{equation}\label{eq14}
	\Omega_r \ll \sqrt{\frac{G\,M_r}{r^3}} = \Omega_{r,\,\rm crit},
\end{equation}
where $\Omega_{r,\,\rm crit}$ is the critical angular velocity at the location point $r$. For a star that rotates rigidly in the radiative zone, we have $\Omega_r=\Omega_{r_1}\approx \Omega_{r_0}$. Since the change in the mass of a star in the radiative zone is less than the change in $r$, we can get $\Omega_{r,\,\rm crit}>\Omega_{r_0,\,\rm crit}$. Thus, $\Omega_r \approx \Omega_{r_0} < \Omega_{r_{0},\,\rm crit} < \Omega_{r, \rm\, crit}$. In summary, in the radiative zone, as long as the rotation velocity of a star does not exceed its critical velocity ($v_{r_0,\rm \, crit} = \sqrt{G\,M_0/r_0}$), Equation (\ref{eq14})  can be considered satisfied.

Specific to our models, in the late of the He flash and the core He burning stages, the radiative zone of the star may maintain the rigid body rotation. During above both phases, based on our model parameters, $r_0\approx10^{12}\,\rm cm$, so $v_{r_0,\rm \, crit}\approx 100\, \rm km\,s^{-1} $. Because of $r_1\approx0.1r_0$, we can infer 
\begin{equation}\label{eq15}
	\Omega_r = \Omega_{r_1} <  \Omega_{r_1, \rm\, crit} \approx 10\,\Omega_{r_{0},\,\rm crit} < \Omega_{r,\,\rm crit}\ \ \ \ \ \ \ (r_2\leqslant r<r_1).
\end{equation}
Therefore, Equation (\ref{eq14}) can be fulfilled when the input velocity approaches the critical value in our models.

In addition, for the rigid body rotation, we require $d\,\Omega_r/d\,r=0$. Although $\Omega_r \ll \Omega_{r,\,\rm crit}$ can be met, $d\,\Omega_r/d\,r=0$ is not necessarily satisfied. Therefore, we still need to add some restrictions in the numerical calculation. The criterion for the rotation of a rigid body we protocolled is
\begin{equation}\label{eq16}
	\left| \frac{\Omega_r-\Omega_{r_1}}{\Omega_{r_1}}\right|<0.1 \ \ \ \ \ \ \ (r_2\leqslant r<r_1).
\end{equation}

\section{Result} \label{sect3}
\subsection{Area for the Rigid Body Rotation} \label{sect31}
Analysing two types of RC star samples, \citet{2023ApJ...944L...5M}  found that only the primary RC stars ($<2.0\,M_{\odot}$) that have experienced the He flash could  exhibit Li-rich characteristics; therefore, we will start thinking about rigid body rotation regions from the He flash stages. We have run a test model and show the angular velocity distribution of the radiative zone in the He flash and the core He burning phases in Figure \ref{af1}. The figure illustrates the outlines of $\Omega_r$ and $X$ between $X=\rm 0.70$ and $r_1$. The regions to the right of the vertical lines satisfy Equation (\ref{eq16}). The results show that during majority of the core He burning stage, the radiative zone between the H burning shell and the convective envelope can satisfy the criterion for rigid body rotation. 
Recently, \citet{2024A&A...681L..20M} found that the differential rotation inside RC stars is locked, and $\Omega_{\rm envelope}\varpropto r^{-2}_0$, this also shows that the RC stars are more likely to satisfy Equation (\ref{eq16}).
Therefore, we will input different equatorial velocities, i.e., $v_{\rm in,\,RC}$, at the beginning of the core He burning, and see Section \ref{sect32}. In addition,  Figure \ref{af2} shows that, when the input velocity $v_{\rm in,\,RC}$ is 1 and $10\,\rm km\,s^{-1}$, the regions conforming to the criterion of the rigid body rotation in the early stages of the RC stars exceed the position of $X=0.70$, i.e., $r_2 < r_{X=0.7}$. Since we are concerned with the region between the H burning shell and the convective envelope, we also consider $r_{X=0.7}$ as the lower boundary of the region of action of $\langle  D_{\rm MC}\rangle$. The trigger condition for this situation is 
\begin{equation}\label{eq17}
	\left| \frac{\Omega_r-\Omega_{r_1}}{\Omega_{r_1}}\right|<0.1 \ \ \ \ \ \ \ (r_2\leqslant r_{X=0.7}\leqslant r<r_1).
\end{equation}
$r_2$, which satisfies Equation (\ref{eq16}) and is greater than $r_{X=0.7}$, is the first choice for the lower boundary. From Figure \ref{af2}, we can find that there is $r_2>r_{X=0.7}$ in most cases. Furthermore, for higher $v_{\rm in,\,RC }$, we find that $r_1-r_2$ is larger. Because stellar core spins faster than its surface envelope, the higher the $v_{\rm in,\,RC }$, the larger the rigid rotation area in the radiative zone will be.

\subsection{Evolution for Lithium Abundance} \label{sect32}
\begin{figure*}[hbt]
	\centering
	\includegraphics[width=18.00cm]{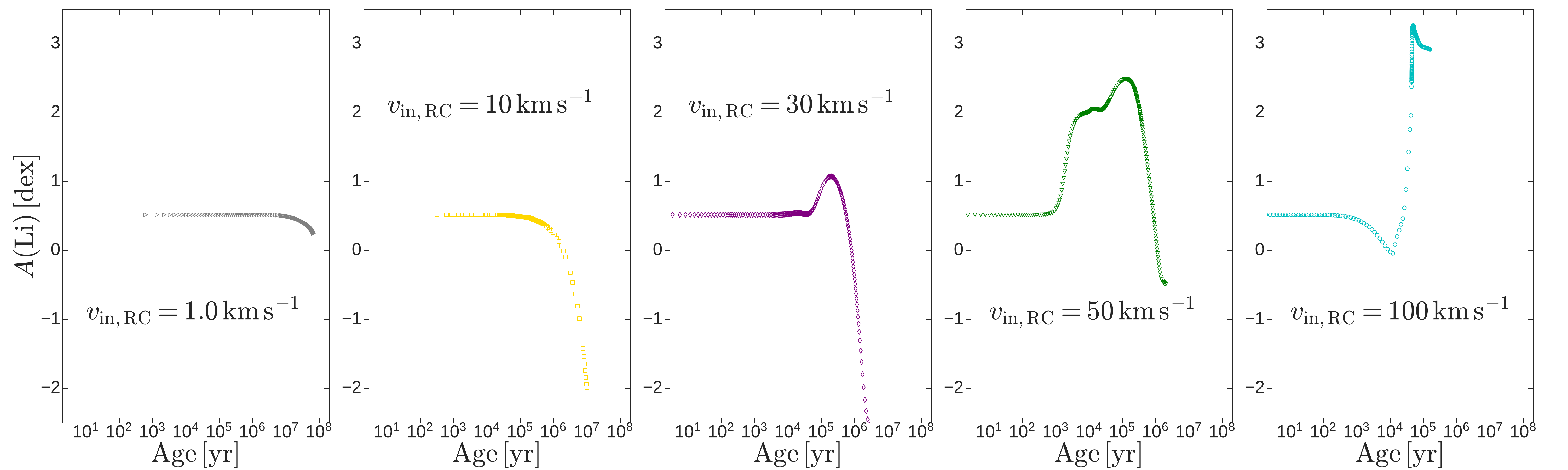}
	\caption{$A(\rm Li)$ vs. the age of RC stars. The projection of the evolution lines on the timeline is the timescale of the meridional circulation. At the start of the core He burning,  $v_{\rm in,\,RC}$ is respectively 1.0, 10, 30, 50, and $100\,\rm km\,s^{-1}$.}
	\label{f1}%
\end{figure*}

Figure \ref{f1} illustrates the evolution trajectory of the surface Li abundance of the RC stars in relation to their ages, and the projection of these lines on the horizontal axis represents $\tau_{\rm MC}$. 
To isolate the effect of meridional circulation, we present in Figure \ref{af3} the evolution of Li abundance predicted by the models with diverse rotation velocities in the absence of meridional circulation (where internal mixing consists of the five rotation instabilities mentioned in Section \ref{sect21}). It can be found that these instability processes have little effect on the evolution of Li abundance; therefore, the results in Figure \ref{f1} fully show the role of meridional circulation.
It can be found that the higher the rotation velocities of a star, the more prone RC star to enrich Li and form the Li-rich RC stars. Notably, higher rotation velocities result in lower $\tau_{\rm MC}$ values, and the timescale of the existence of the Li-rich giants is about $10^6\rm \,yr$, i.e., the Li-rich RC stars are relatively young. This is line with the asteroseismic observed evidences of \citet{2021ApJ...913L...4S}. In addition, they indicated a rapidly decay of about 3 orders of magnitude after Li enrichment may occur, which is similar to the evolutionary behavior in timescale $\tau_{\rm MC}$ predicted by our models.

The Li abundance forecasted by our models is $\rm 0.52\,dex$ at the zero age of the  core He burning \citep[which corresponds to the lower limit of Li abundance distribution for the RC stars; see][]{2021ApJ...919L...3Z}, and the maximum Li abundance increment can be close to $3.0\,\rm dex$ in the case of $v_{\rm in,\,RC}=100\,\rm km\,s^{-1}$. Consequently, our results demonstrate that the mixing process driven by meridional circulation can lead to the formation of super Li-rich RC stars with \(A(\rm Li) > 3.3\,\rm dex\) when the rotation velocity close to the critical value. In addition, the generation of Li-rich RC stars requires maintaining a rotation velocity on the order of $10\,\rm km\,s^{-1}$. 

\begin{figure*}[hbt]
	\centering
	\includegraphics[width=16.00cm]{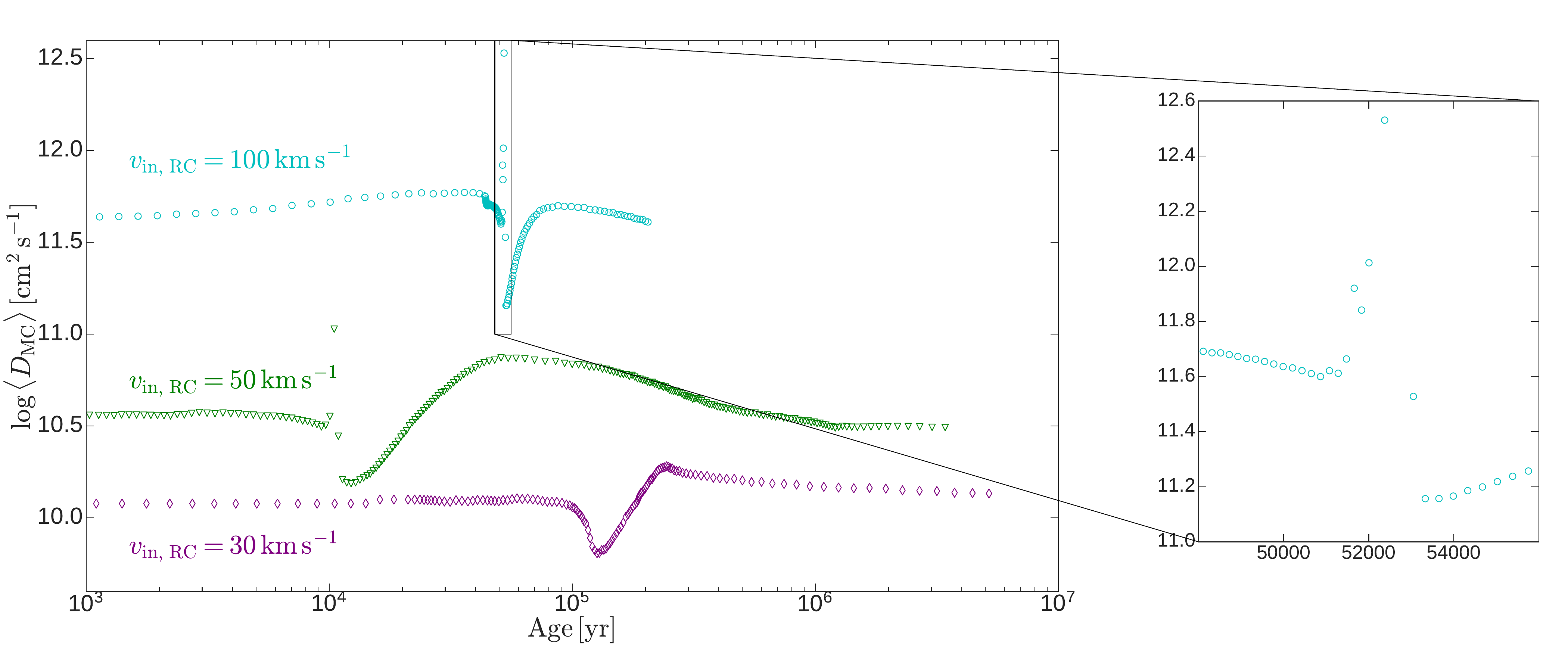}
	\caption{Evolution of diffusion coefficient with the age of the RC stars. The input equatorial velocities corresponding to the three evolutionary trajectories are 30, 50, and $\rm 100\,km\,s^{-1}$, respectively.}
	\label{f2}%
\end{figure*}

For the cases of low rotation velocity ($v_{\rm in,\,RC} = 1$ and $10\,\rm km\,s^{-1}$), because $\langle D_{\rm MC} \rangle$ is relatively low, $A(\rm Li)$ will not increase but decrease. At a rotation velocity of $30\,\rm km\,s^{-1}$, the RC stars show Li enrichment behavior first of all, but then also reveal Li depletion process. At higher rates, the evolution of $A(\rm Li)$ get a little more complicated. 

Above results can be attributed to the magnitude change and evolution of $\langle D_{\rm MC}\rangle$. 
When $v_{\rm in,\,RC} = 1$ and $10\,\rm km\,s^{-1}$, their $\langle D_{\rm MC}\rangle$ differs by nearly two orders of magnitude. $\langle D_{\rm MC}\rangle$ is very small ($\sim 10^7\,\rm cm^2\,s^{-1}$) for $v_{\rm in,\,RC} = 1\,\rm km\,s^{-1}$, limiting the transport of material, so the abundance remains basically unchanged. $\langle D_{\rm MC}\rangle$ is slightly higher and about $10^9\,\rm cm^2\,s^{-1}$ when $v_{\rm in,\,RC} = 10\,\rm km\,s^{-1}$, the material transport becomes significant, but the inflow of beryllium in the convective envelope does not exceed the outflow of Li, so the Li depletion will be more obvious. $\langle D_{\rm MC}\rangle$ is closely related to stellar parameters, and the models corresponding to Figure \ref{f1} only varies in input velocities, so we only display the evolution of $\langle D_{\rm MC}\rangle$ over time in the cases of $v_{\rm in, \, RC}=30,\ 50,\ \rm and\ 100\,\rm km\,s^{-1}$ in Figure \ref{f2}. For $v_{\rm in,\,RC} = 100\,\rm km\,s^{-1}$, the results show that $\langle D_{\rm MC}\rangle$ shows a downward trend as a whole when the age of the RC star is $\sim 10^5\,\rm yr$, and therefore $A(\rm Li)$ decreases. Similar case occurs for $v_{\rm in,\,RC} = 50\,\rm km\,s^{-1}$ and $v_{\rm in,\,RC} = 30\,\rm km\,s^{-1}$, but the onset of Li depletion is later. It can be noted that the maximum degree of Li enrichment predicted by our models occurs with diffusion coefficient of about \(10^{11.5}\,\rm cm^2\,s^{-1}\), which is consistent with the result corresponding to the maximum Li enrichment of \citet{2004ApJ...612.1081D}.

It is observable in Figure \ref{f1} that Li depletion occurs near an age of \(10^4\,\rm yr\)  for the case where \(v_{\rm in,\,RC} = 100\,\rm km\,s^{-1}\). This depletion is attributed to the small scale of the rigid body part of the radiative zone at this stage, as depicted in the first subfigure of the lower panel of Figure \ref{af2}, which prevents the transfer of beryllium into the convective envelope. The primary reason for this phenomenon is that we assign different rotation velocities at the zero age of the core He burning, and when the rotation velocity is high, the star requires additional time to adjust its structure, resulting in a small rigid region initially. As the stellar structure adjusts, the rigid region expands gradually. The last subfigure of Figure \ref{f1} indicates that the star takes approximately \(10^4\,\rm yr\) to adjust its structure to form the rigid region we needed when the input velocity is near the critical velocity.

\citet{2004ApJ...612.1081D} highlighted that high angular velocities are sufficient conditions for the formation of high Li giants, and they utilized the interaction between close binaries as a way to accelerate the target star. This acceleration process will be not last long \citep{2019ApJ...880..125C}. While there is uncertainty regarding whether the RC stars are indeed spinning up, we can reasonably assume that they maintain a comparable velocity throughout the core He burning phase, given that the timescale of this phase significantly exceeds the duration of the aforementioned acceleration. Consequently, we can introduce a range of velocities into our models during the He flash, ensuring that the RC stars do not adjust the structure due to apparently spin up. 

\begin{figure*}[hbt]
	\centering
	\includegraphics[width=18.00cm]{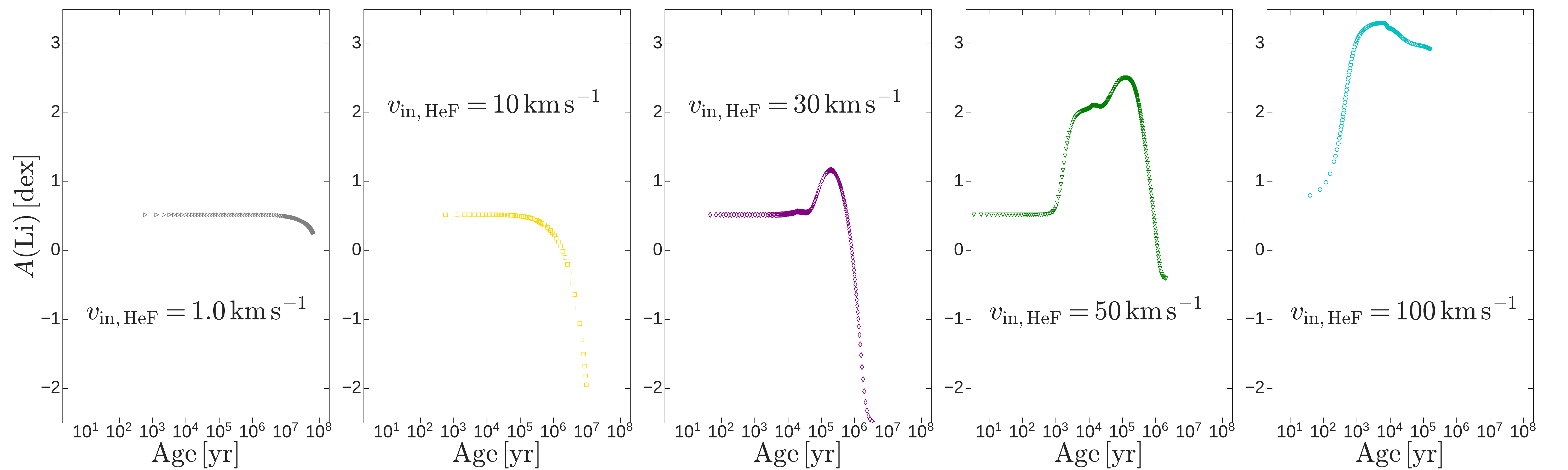}
	\caption{Similar to Figure \ref{f1}, but the input velocity is $v_{\rm in,\,HeF}$.}
	\label{f3}%
\end{figure*}
The star will contract rapidly in the early phases of the He flash. To ensure that the velocity of the core He burning phase does not exceed the critical velocity of $
\sim 100\,\rm km\,s^{-1}$, we can input the different velocities, $v_{\rm in,\,HeF}$, at time $S$ (see Figure \ref{af4}). After time $S$, the subsequent true rotation velocity will be less than $v_{\rm in,\,HeF}$ since the radius of the star is increasing.
Figure \ref{f3} presents the evolution information of the corresponding cases. An evolutionary behavior of Li abundance similar to that shown in Figure \ref{f1} can be seen, and in addition, the Li depletion that occurs at about \(10^4\,\rm yr\) in the case of $v_{\rm in,\,RC}\sim 100\,\rm km\,s^{-1}$ is replaced by an increase in Li abundance after the structural adjustment effect caused by rotation is eliminated.

As can be seen from Figure \ref{af1}, at low velocity, the target radiative zone in the late stage of the He flash does not fully satisfy Equation (\ref{eq16}). With varying $v_{\rm in,\,HeF}$, we show in Figure \ref{af5}
the contours of $\Omega_r$ and $X$ of the radiative zone in the He flash phase after time $S$. It is found that when the rotation velocity is high, the radiative zone in the later He flash can satisfy Equation (\ref{eq16}). Thus, the Li enrichment can also occur in the late stage of the He flash, which leads to the formation of Li-rich RC stars since its evolution time is $\sim 10^5\,\rm yr$ ($< \tau_{\rm MC}$).

\section{Discussion}\label{sect4}
\subsection{Diffusion Coefficient}
Based on the description of the radial mixing of meridional circulation to the element transfer in \citet{1979ApJ...229..624S}, we give the expression of the diffusion coefficient from the relationship between the diffusion coefficient and the mass change (see Equation (\ref{eq3})). Unlike the perspective of the dimension, the usual way of dealing with the diffusion coefficient is the product of velocity and distance. For example, the diffusion coefficient for mixing process driven by internal gravity waves is calculated as follow: calculating the velocity field driven by the disturbance of the convective boundaries first, and then considering the displacement change in the irreversible oscillate process of the fluid element \citep[see, e.g.,][]{1994A&A...281..421M}. For the macroscopic meridional circulation, the position change caused by the change of velocity is considered, that is, the velocity scale height ($H_v=| d\,r / d\,\text{ln}\, v_{\rm MC} |$) is introduced \citep[see][]{1978ApJ...220..279E}. Similarly, if we replace the $\Delta r$ in Equation (\ref{eq3}) with $H_v$ (or the pressure scale height $H_P$). Since the magnitude of $H_v$ is $\sim 10^9\,\rm cm$\footnote[3]{As can be seen from Equations (\ref{eq6}) and (\ref{eq7}), the change in velocity is about 3 orders of magnitude in the radiative zone ($v_{\rm MC}\varpropto r^3$). The pressure changes in the corresponding region are about 3-4 orders of magnitude. Because of sharing similar definition, $H_v$ is actually slightly greater than $H_P$ ($\sim 10^9\,\rm cm$). Here, we take $H_v$ roughly as $10^9\,\rm cm$.}, the length term of Equation (\ref{eq3}) has a magnitude of $\sim 10^{13}\,\rm cm$. 
In the case of $v_{\rm in,\,RC}=50\,\rm km\,s^{-1}$, $\langle v_{\rm MC}\rangle$ is about $10^{-2}\,\rm cm\,s^{-1}$, so the diffusion coefficient is $\sim 10^{10-12}\,\rm cm^2\,s^{-1}$ in the radiative zone. This is similar to what our mechanism predicts for $\langle D_{\rm MC}\rangle$.

On the other hand, since diffusion is a microscopic process, the diffusion coefficient can also be obtained from the diffusion equation
\begin{equation}
	\rho \frac{dX_i}{dt}=\rho \frac{\partial X_i}{\partial t} +\rho\, \vec{u}\cdot \nabla X_i = \nabla \cdot (\rho\, D\, \nabla X_i),
\end{equation}
here $X_i$ is the abundance of element $i$ and $\vec{u}$ is the vector velocity. Then, \citet{1992A&A...253..173C} and \citet{1992A&A...265..115Z} gave the expression of diffusion coefficient for meridional circulation without ignoring the baroclinic instability
\begin{equation}
	D_{\rm eff} = \frac{(r\, v_{\rm MC})^2}{30 D_{\rm h}},
\end{equation} 
where $D_{\rm h}$ is the horizontal turbulent diffusion coefficient. Due to mathematical difficulties, \citet{2000ApJ...528..368H} pointed out that the above process is applicable in the main sequence stars, although \citet{2000MNRAS.316..395D} extended this way of dealing with diffusion to giants. 
Without considering horizontal turbulence, \citet{2009pfer.book.....M} made a rough estimate and got $D_{\rm MC}\sim r\,v_{\rm MC}$. It is $\sim 10^{9}\,\rm cm^2\,s^{-1}$ when $v_{\rm in,\,RC}=50\,\rm km\,s^{-1}$. This is slightly lower than the mean effect of predicted our mechanism, which will weaken the Li enrichment effect predicted by our models. 

In short, although there are many ways of dealing with meridional circulation, in general, they do not differ greatly in terms of material transport. In rotating stars, the key factor in determining Li enrichment is the rotation velocity. The efficiency of mixing process can be significantly improved with the increase of rotation velocity.

To be involved in mixing in a radiative zone, for the thermal mixing in the radiative zone, rising and sinking fluid elements must have enough time to exchange heat with their surroundings, which is done by radiative diffusion. Therefore, the radiative diffusivity $D_{\rm rad}$\footnote[4]{$D_{\rm rad}=\frac{c}{3\,\kappa\,\rho}$, where $c$ is the light speed in vacuum and $\kappa$ is the opacity.} sets up the upper limit for a thermal diffusion coefficient. Based on the RC model parameters, we can expect $\langle D_{\rm MC}\rangle$ to exceed $D_{\rm rad}$ in the cases of rotation velocities with 50 and $100\,\rm km\,s^{-1}$. However,
it should be noted that meridional circulation is a macroscopic large scale fluid motion rather than a microscopic thermal diffusion process \citep[e.g.][]{1992A&A...253..173C}. While we model it as a diffusion process to assess its mixing effects, this approach is a simplification. 

\subsection{Rotation Velocity and Lithium Abundance}
In rotating stars, meridional circulation is a process that must be considered, so our scenario is general and does not depend very much on the parameters of a star, which may be regard as a universal Li enrichment process. But certain phenomena still need to be explained further.
\begin{enumerate}
	\item \textit{\textsf{low-velocity Li-rich RC stars}}. As shown in Figure \ref{f1}, stars with a specific rotation velocity usually have a Li abundance distribution, and the higher the rotation velocity, the larger the distribution with $A(\rm Li)>\,0.5\,dex$ is. Therefore, we cannot make a good summary of the exact relationship between Li abundance and velocity, but it is clear that there is a positive correlation between Li enrichment and velocity. This aligns with the results of \citet{2023AJ....166...60T}, they also showed that the actual correlation between Li abundance and rotation rate is weak. In other words, the low-velocity Li-rich RC stars deviate from what our model predicts. Here, we give two possible explanations. On the one hand, observations are usually given the projected component of the equatorial velocity ($v\,sin i$), whereas our models only use the equatorial velocity. Because each velocity corresponds to the distribution characteristics of Li abundance. Thus, the high Li abundance and the Li abundance distribution of some low-velocity RC stars (lower $v\,sin i$) may be explained by our models as well. On the other hand, when a star evolves from the RGB to the RC phase, \citet{2012A&A...548A..10M} found that the rotation velocity of the stellar core noticeably decreases. They suggested this could be due to the expansion of the non-degenerate He burning core, leading to the deceleration of the stellar core. Later, \citet{2024A&A...681L..20M} revealed that the differential rotation in RC stars is locked. Therefore, a decrease in the rotation velocity of the stellar core may also cause the surface envelope to slow down during the core He burning, resulting in a decrease in the overall stellar rotation velocity. When our mechanism is adopted in the later He flash phase (see Section \ref{sect32}), the rotation velocity of the star will not decrease until the star initiates central He burning. The evolution time in this late He flash phase is \(\sim 10^5\,\text{yr}\), which is close to the Li enrichment timescale shown in Figure \ref{f1}, allowing the star to achieve Li enrichment first, and then spin down. Consequently, this may also promote the formation of low-velocity Li-rich RC stars.
	
	\item \textit{\textsf{Around 1\% Li-rich giants}}. At present, Li-rich giants are very few. Our models present that when the equatorial rotation velocity exceeds $\rm 30\,\rm km\,s^{-1}$, it is possible for a RC star to have Li-rich characteristics. Current observations suggest that low-velocity giants are more common. Recently, \citet{2024MNRAS.528.3232P} found that the high-velocity giant candidates ($v\,sini>10\,\rm km\,s^{-1}$, and about 10\%) are present in the APOGEE DR16 survey, while only 1\% are present in the \textit{Kepler} control sample.
	In addition, the samples of \citet{2021A&A...651A..84M} (from the Gaia-ESO survey and \citet{1999A&AS..139..433D}) also show the scarcity of the high-velocity giants. In a subset of the APOGEE survey, for the RC stars, the proportion of the  high-velocity objects is only about 1/30 \citep{2023AJ....166...60T}. These observations will strictly restrict the formation rate of Li-rich giants. If we assume that an equatorial velocity of $\rm 30\,\rm km\,s^{-1}$ corresponds to $v\,sini=10\,\rm km\,s^{-1}$, then our results predict a generation rate of Li-rich giants in the range of $\sim 1\%-10\%$. Considering the average effect and making a rough estimate, $\langle v\,sini\rangle>10\,\rm km\,s^{-1}$ necessitates a mean equatorial velocity of over $\rm 16\,\rm km\,s^{-1}$. This will reduce the formation rate of Li-rich giants and could form $\sim 1\%$ Li-rich giants. In general, from the perspective of the distribution of rotation velocities, our Li enrichment mechanism predicts the scarcity of Li-rich giants.
	
	\item \textit{\textsf{Li-rich RGB stars.}} The self-consistent model by \citet{2006A&A...453..261P} indicates that meridional circulation cannot drive sufficient mixing to achieve effective Li enrichment for RGB stars. Here, applying our mechanism to the RGB stars, due to limitations imposed by two factors, our models will deplete Li in the RGB stars. The first factor is the rotation velocity; the stellar radius increases during the first ascent of the giant branch, thus its velocity does not become higher. The second factor is the rapid expansion of the star at this stage, resulting in a relatively small radiation zone within the star that satisfies Equation (\ref{eq16}). A lower velocity would predict the Li depletion, and a narrow uniformly rotating radiation zone would also cause parallel Li behaviour (see Figure \ref{f1}). The two causes are expected to lead to the Li depletion of the RGB stars. This conforms to the primary behaviour of Li in the RGB \citep{2020NatAs...4.1059K}. However, the Li-rich RGB stars cannot be deciphered by our models, moreover, the present most Li-rich RGB star has a $A(\rm Li)$ of $\sim 6.0\,\rm dex$ \citep{2024ApJ...973..125K}. This will urge us to explore other Li enrichment processes.
\end{enumerate}

Although more comprehensive statistics for $A(\text{Li})-v\,sini$ are not yet available, the above discussions indicate that our mechanism can well be suited to explain the Li phenomenon of RC stars.

\section{Conclusion}\label{sect5}
In this study, we investigate the Li enrichment mechanism in rotating RC stars. We consider meridional circulation within the star and treat its radial mixing as a diffusion process. The diffusion coefficient is derived from the mass change in the material transfer process, applied to the core He burning stage, and the Li enrichment effect is explored at various rotation velocities. Our results reveal a strong correlation between the degree of Li enrichment and rotation velocity: the higher the rotation velocity, the more significant the Li enrichment, with a maximum increment approaching \(3.0\,\rm dex\). Thus, our mechanism predicts the formation of Li-rich and potentially super Li-rich RC stars, which are typically characterized by higher equatorial velocities. The Li enrichment is likely to occur during the late He flash and early stages of the core He  burning. Therefore, our models suggest that Li anomalous giants are relatively young, at approximately \(10^6\,\rm yr\).

\begin{acknowledgments}

We appreciate the insightful comments and constructive help from the anonymous reviewer. The work is supported by National Natural Science Foundation of China (grant Nos: 11973079, 12288102, 12133011, 11973052, 12022304, 12090040, 12090044, 12173080, 12273104, and 12373036) and the National Key R\&D Program of China Nos. 2021YFA1600400, 2021YFA1600402. This research is funded by a grant from the National Basic Science Center Project of China (grant No. 12288102), the Natural Science Foundation of Yunnan Province (grant No. 202201AT070158), and the Yunnan Fundamental Research Projects (grant No. 202401AS070045). H.-L. Y. acknowledges support from the Youth  Innovation Promotion Association of the CAS and the NAOC Nebula Talents Program. J.-H. Z. acknowledges support from NSFC grant No.12103063 and from China Postdoctoral Science Foundation funded project (grant No. 2020M680672).
\end{acknowledgments}




\appendix
\setcounter{figure}{0} 
\section{Supplementary graphs for supporting the result analyses}
\begin{figure*}[hbt]
	\centering
	\figurenum{A1}
	\includegraphics[width=18.00cm]{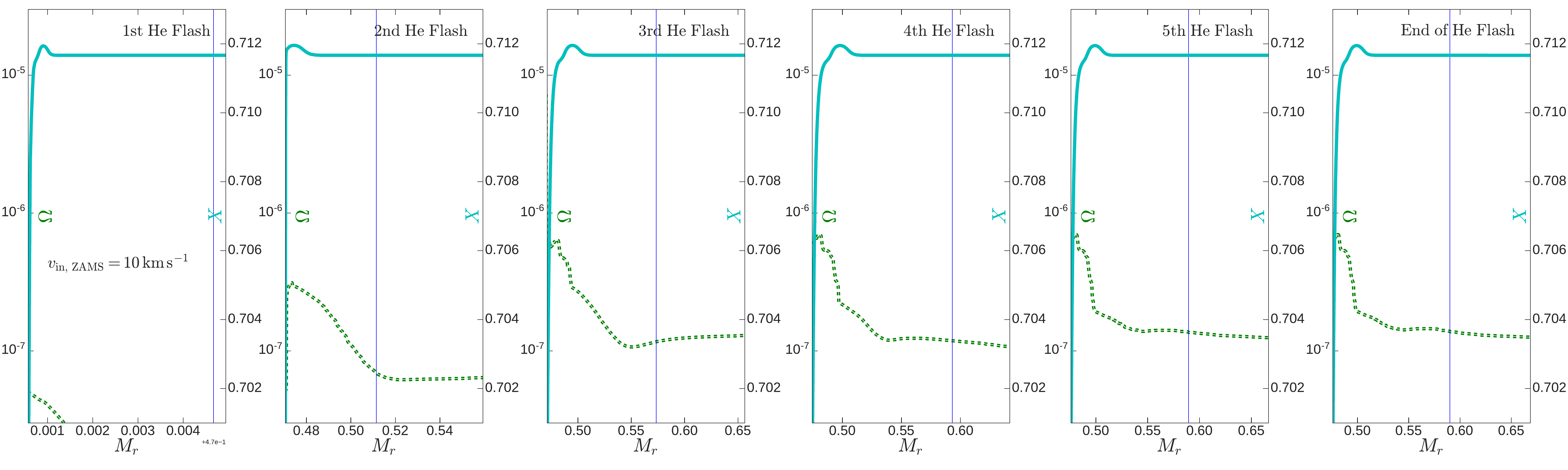}
	\includegraphics[width=18.00cm]{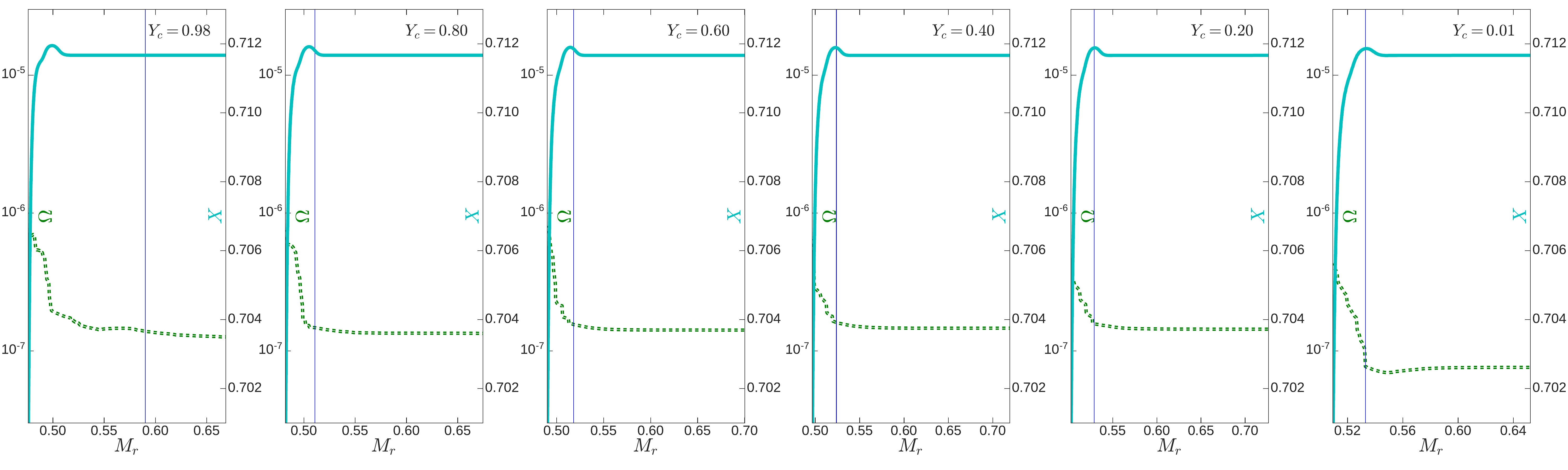}
	\caption{The angular velocity distribution and mass fraction of H in the He flashes and core He burning phases. Stellar parameters are $1.2\,M_{\odot}$, $Z=0.017$, $v_{\rm in,\,ZAMS}=10\,\rm  km\,s^{-1}$, and $A(\rm Li)_{\rm in,\,ZAMS}=3.3\,dex$. Here, no input velocity $v_{\rm in,\,RC}$. The two ends of the horizontal axis are $X=0.70$ and the bottom of the convective envelope. The vertical lines represent $\Omega_r=(1+|0.1|)\,\Omega_{r_1}$, then the right side of which is regarded as the rigid body. $Y_c$ is the mass fraction of central He. Note that the figure does not include the effect of meridional circulation.}
	\label{af1}%
\end{figure*}

\begin{figure*}
	\centering
	\figurenum{A2}
	\includegraphics[width=18.00cm]{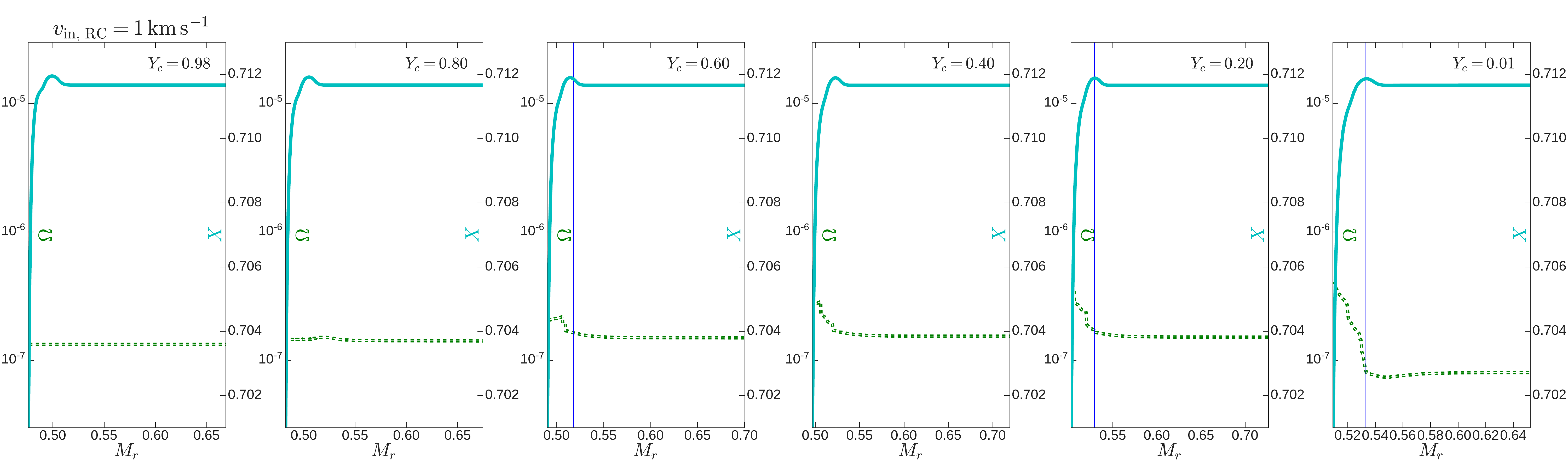}
	\includegraphics[width=18.00cm]{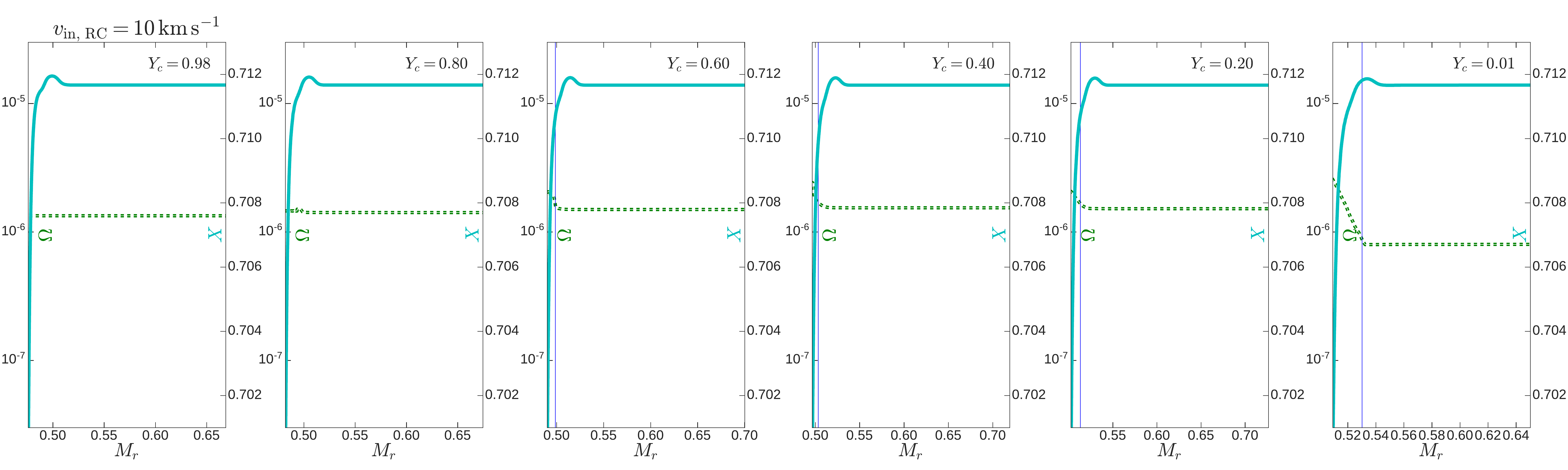}
	\includegraphics[width=18.00cm]{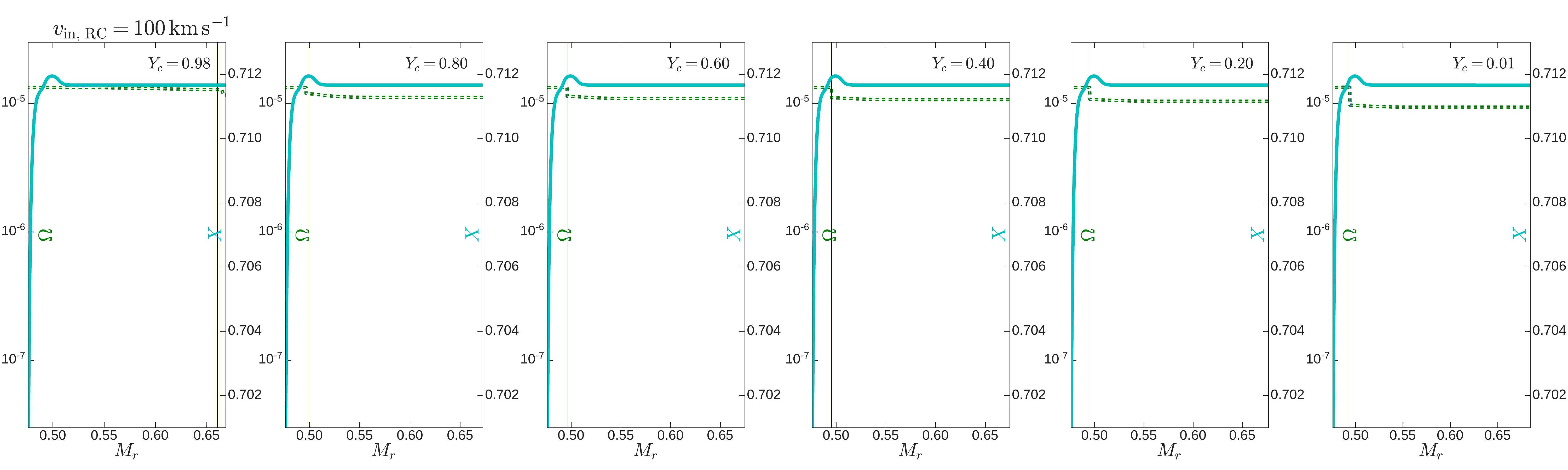}
	\caption{Similar to $\rm Figure\,\ref{af1}$, but only showing the distributions of $\Omega_r$ and $X$ in the core He burning stage. The input velocity of the three panels $v_{\rm in,\,RC}$ is 1, 10, and $100\,\rm km\,s^{-1}$, respectively.}
	\label{af2}%
\end{figure*}

\begin{figure*}[hbt]
	\centering
	\figurenum{A3}
	\includegraphics[width=18.00cm]{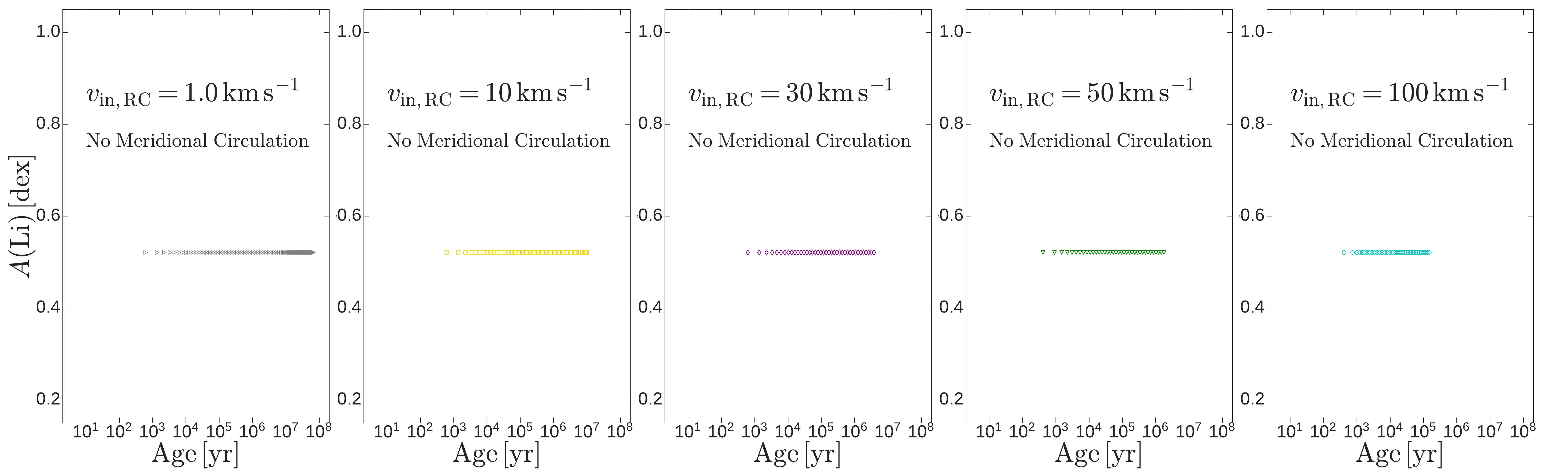} 
	\caption{Similar to Figure \ref{f1}, but the transport process of material of meridional circulation is excluded.}
	\label{af3}%
\end{figure*}

\begin{figure*}[hbt]
	\centering
	\figurenum{A4}
	\includegraphics[width=12.00cm]{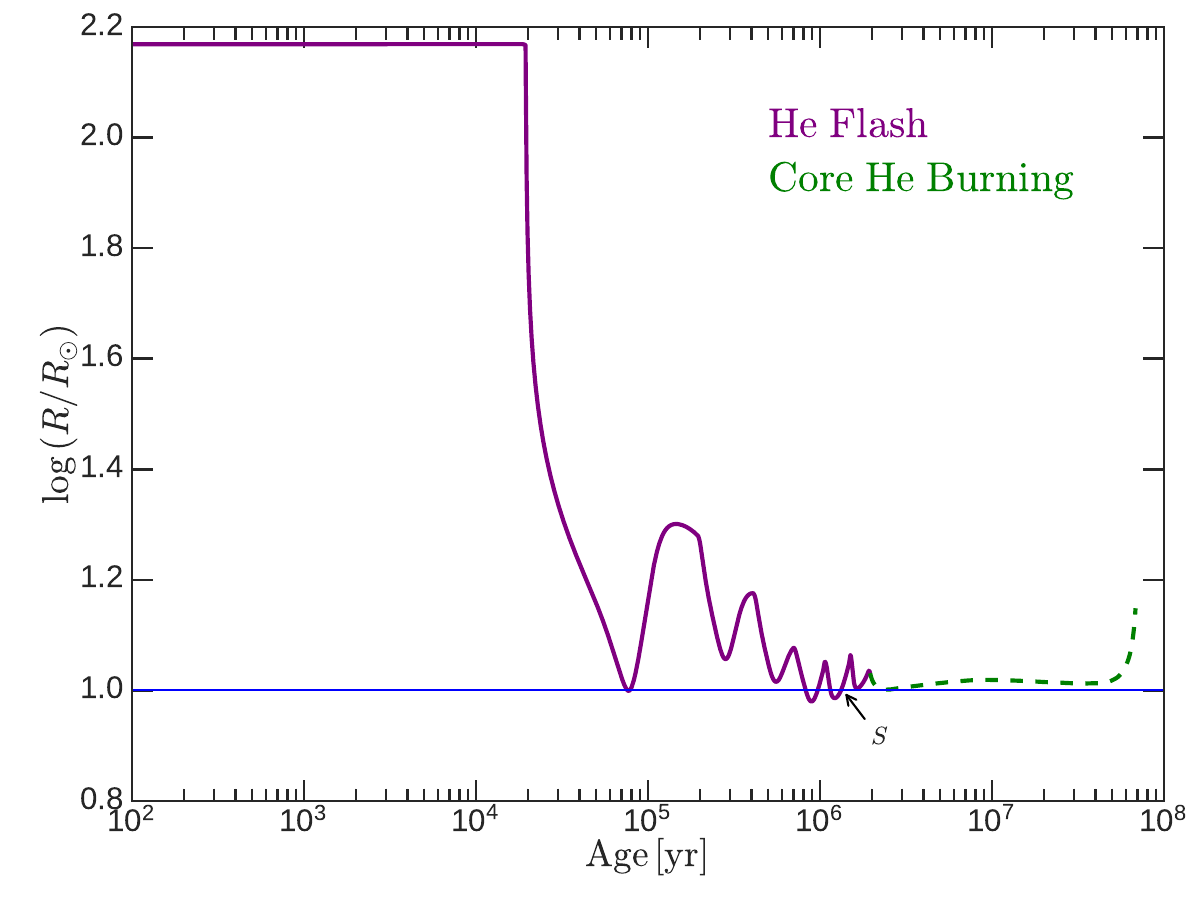}
	\caption{Evolution for stellar radius with age in the He burning phase. We take the time of the He flash stage corresponding to the intersection $S$ as the input moment of $v_{\rm in,\,HeF}$. The radius of the star at time $S$ is the minimum value for the entire core He burning phase.}
	\label{af4}%
\end{figure*}

\begin{figure*}
	\centering
	\figurenum{A5}
	\includegraphics[width=18.00cm]{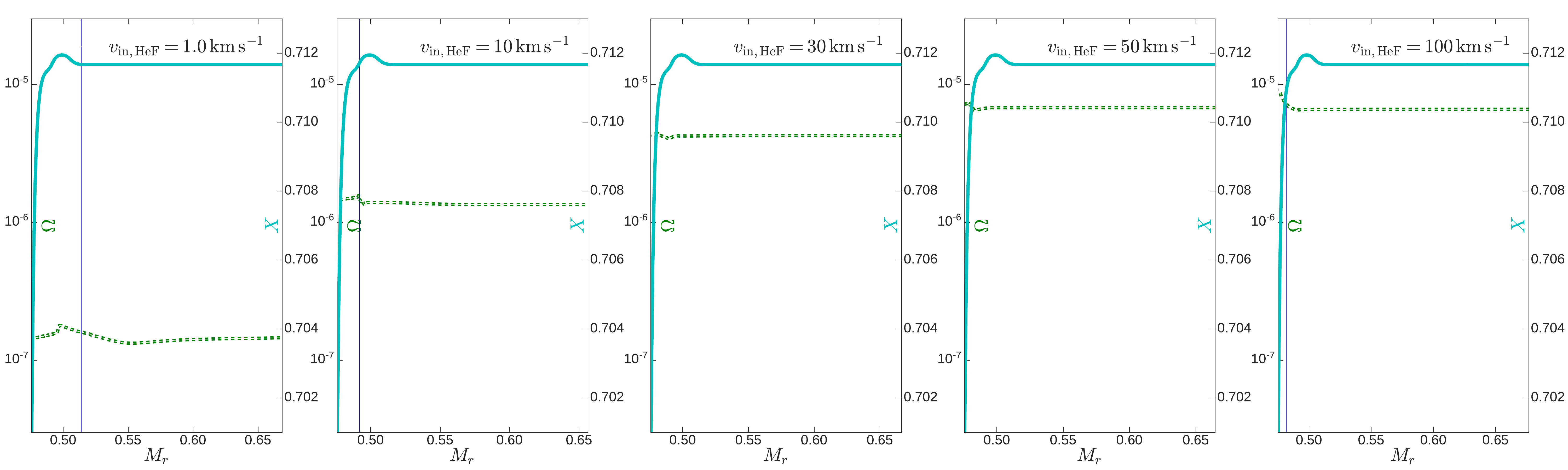}
	\caption{Similar to Figure \ref{af1}, but only showing the distributions of $\Omega_r$ and $X$ in the late stage of He flash (after the time $S$ in Figure \ref{af4}). The input velocity $v_{\rm in,\,HeF}$ is 1, 10, 30, 50, and $100\,\rm km\,s^{-1}$, respectively.}
	\label{af5}%
\end{figure*}

\FloatBarrier




\end{document}